\documentstyle[12pt]{article}

\advance\voffset by -0.5cm
\advance\hoffset by -1.5cm
\input amssym.def
\input amssym.tex
\def\be{\begin{equation}}
\def\ee{\end{equation}}

\def\Zop{{\Bbb Z}}
\def\Rop{{\Bbb R}}

\def\pmb#1{\setbox0=\hbox{#1}%
 \kern-.025em\copy0\kern-\wd0
 \kern.05em\copy0\kern-\wd0
 \kern-.025em\raise.0433em\box0 }

\def\jbar{\overline{j}}
\def\lambdabar{\overline{\lambda}}

\def\B{{\cal B}}
\def\M{{\cal M}}

\def\Z{{\cal Z}}
\def\H{{\cal H}}
\def\W{{\cal W}}

\def\bW{{\overline{\cal W}}}
\def\bPhi{{\overline{\Phi}}}
\def\bGamma{{\overline{\Gamma}}}
\def\tg{{\tilde g}}
\def\lg{{\frak g}}
\def\lh{{\frak h}}
\def\tG{{\widetilde G}}
\def\tu{{\tilde u}}
\def\tv{{\tilde v}}

\def\tL{{\tilde L}}

\def\3{\ss}
\def\sq{\hbox{\rlap{$\sqcap$}$\sqcup$}}
\def\qed{\ifmmode\sq\else{\unskip\nobreak\hfil
\penalty50\hskip1em\null\nobreak\hfil\sq
\parfillskip=0pt\finalhyphendemerits=0\endgraf}\fi}
\def\half {\frac{1}{2}}

\def\noi {\noindent}

\def\bbbr {{\rm I\!R}}

\def\bbbone {{\mathchoice {{\rm 1\mskip-4mu l}} {{\rm 1\mskip-4mu l}}
{{\rm 1\mskip-4.5mu l}} {{\rm 1\mskip-5mu l}}}}
\def\bbbc{{\mathchoice {\setbox0=\hbox{$\displaystyle\rm C$}\hbox{\hbox
to0pt{\kern0.4\wd0\vrule height0.9\ht0\hss}\box0}}
{\setbox0=\hbox{$\textstyle\rm C$}\hbox{\hbox
to0pt{\kern0.4\wd0\vrule height0.9\ht0\hss}\box0}}
{\setbox0=\hbox{$\scriptstyle\rm C$}\hbox{\hbox
to0pt{\kern0.4\wd0\vrule height0.9\ht0\hss}\box0}}
{\setbox0=\hbox{$\scriptscriptstyle\rm C$}\hbox{\hbox
to0pt{\kern0.4\wd0\vrule height0.9\ht0\hss}\box0}}}}

\begin{document}
\thispagestyle{empty}
\begin{flushright}
DAMTP-95-46
\end{flushright}
\vspace{2.0cm}

\begin{center}

{\Large {\bf WZW models of general simple groups}}
\vspace{2.0cm}

{\large Matthias R. Gaberdiel}
\footnote{e-mail: M.R.Gaberdiel@damtp.cam.ac.uk} \\
{Department of Applied Mathematics and Theoretical
Physics\\
University of Cambridge, Silver Street \\
Cambridge, CB3 9EW, U.\ K.\ }
\vspace{0.5cm}

August 1995
\vspace{0.5cm}

{\bf Abstract}
\end{center}

{\leftskip=1.8truecm
\rightskip=1.8truecm

It is shown that a WZW model corresponding to a general simple group
possesses in general different quantisations which are parametrised by
$\mbox{Hom}(\pi_1(G),\mbox{Hom}(\pi_1(G),U(1)))$. The quantum theories
are generically neither monodromy nor modular invariant, but all the
modular invariant theories of Felder {\it et.al.} are contained among
them.

A formula for the transformation of the Sugawara expression for $L_0$
under conjugation with respect to non-contractible loops in $LG$ is
derived. This formula is then used to analyse the monodromy properties
of the various quantisations. It turns out that for
$\pi_1(G)\cong \Zop_N$, with $N$ even, there are $2$ monodromy
invariant theories, one of which is modular invariant, and for
$\pi_1(G)\cong \Zop_2\times\Zop_2$ there are $8$ monodromy invariant
theories, two of which are modular invariant. A few specific examples
are worked out in detail to illustrate the results.

}


\section{Introduction}
\renewcommand{\theequation}{1.\arabic{equation}}
\setcounter{equation}{0}

Among the various conformal field theories, the Wess-Zumino-Witten
(WZW) models \cite{Wit,KZ,GW} take a somewhat special position. First
of all, due to the vast knowledge about the representation theory of
the affine Kac-Moody algebras \cite{PS,Kac,GO}, their mathematical
structure is well understood from an algebraic point of view. On the
other hand the models possess a formulation in terms of an action, and
thus more conventional techniques may be used. In addition, even though
rather special, a very large class of conformal field theories can be
constructed from WZW models, using the coset construction \cite{GKO}.
\smallskip

\noi Most of the work which has been done on WZW models has only taken
into account the local structure of the underlying (target space)
group, ignoring global topological effects. In this paper we shall try
to understand some of these global issues. In particular we shall be
interested in WZW models of groups which are not simply-connected.

\noi Whereas the algebraic approach to WZW models is more powerful for
local considerations, the formulation in terms of an action allows a
discussion of the global issues, and we shall thus take it as our
starting point. We shall explain how the theory can be quantised, and
show how the spectrum of the corresponding quantum theories can be
described algebraically. We shall then analyse how many inequivalent
quantisations exist and exhibit them explicitly. We find that the
various quantisations are parametrised by
$\mbox{Hom}(\pi_1(G),\mbox{Hom}(\pi_1(G),U(1)))$, where $\pi_1(G)$ is
the fundamental group of the group $G$ under consideration.

\noi Having given the various quantisations explicitly, we shall study
some of their properties in detail. In particular, we shall analyse
the behaviour of the fields under monodromy, {\it i.e.} the analytic
continuation of one field about another one, and we shall show that,
for general quantisations, the operator product transforms with
respect to a (non-trivial) one-dimensional representation. This
implies in particular that the amplitudes are only defined on some
covering space. The appearance of an `anyonic' representation is quite
generic for two-dimensional quantum field theories (see {\it e.g.}
\cite{FM,Fre}), and it suggests that the theory is genuinely
braided. In a string theory inspired context, such theories have
traditionally been excluded; however, from the point of view of
Euclidean conformal field theory, this restriction does not seem to be
justified.

\noi The class of theories for which all operator products are
invariant under monodromy, the {\it monodromy invariant theories}, are
of special interest, not least from the traditional point of view of
Euclidean conformal field theory. We shall show that, depending on the
structure of the fundamental group of $G$, one, two or eight of the
different quantisations are monodromy invariant. In more detail, for
$\pi_1(G)\cong \Zop_N$, there exists one monodromy invariant theory,
unless $N$ is even, in which case there are two. One of the monodromy
invariant theories is the known modular invariant theory
\cite{Ber,FGK2}, and the other (for even $N$) is not modular
invariant.  For $\pi_1(G)\cong \Zop_2\times\Zop_2$, there are eight
monodromy invariant quantisations, two of which are the modular
invariant theories of Felder {\it et.al.} \cite{FGK2}, and the other
six are not modular invariant.

\noi The essential calculational tool for the analysis of the
monodromy is a formula for the adjoint action of non-contractible
loops in $LG$ on the generator $L_0$ which we derive.  This
formula may have some interest in its own right as the action of the
non-contractible loops in $LG$ on $\Rop\oplus l\lg\oplus\Rop$, the Lie
algebra of $\Pi\triangleright \tL\tG$ \cite{PS}, a priori involves a
choice; here this choice is fixed by identifying the generator of the
rigid rotations, $L_0$, with the Sugawara expression which is quadratic
in the Lie algebra of $\tL \tG$, $\hat{\lg}\cong l\lg\oplus\Rop$.
\medskip

\noi The paper is organised as follows. We start in section~2 by
studying the quantisation of the theory defined by an action, and
explain how the different quantisations arise, giving explicit
formulae for the spectrum of all possible quantisations. In section~3,
we determine the adjoint action of non-contractible loops in $LG$ on
the generator $L_0$, and show that all quantisations transform with
respect to a one-dimensional representation under monodromy. We then
analyse in section~4 which quantisations are monodromy invariant.  In
section~5, some examples for the additional (monodromy invariant)
theories are exhibited in detail, and section~6 contains a few
concluding remarks. In the appendix, we recall some of the less widely
known facts about the affine Weyl group, and give a new geometrical
proof for an old observation of Olive and Turok \cite{OT} about the
symmetries of the affine Dynkin diagram.

\section{The different quantisations}
\renewcommand{\theequation}{2.\arabic{equation}}
\setcounter{equation}{0}

We want to consider the theory defined by the WZW action
\cite{Wit,FGK1,FGK2}
\be
\label{action}
{\cal S}[g] = -\frac{k}{4 \pi} \int_{\M}
\left\langle g^{-1} \partial g, g^{-1} \bar{\partial} g \right\rangle
- \frac{k}{24 \pi} \int_{\B}
\left\langle \tg^{-1} d \tg,
[\tg^{-1} d \tg, \tg^{-1} d \tg] \right\rangle \,.
\ee
Here the field $g:{\cal M} \rightarrow G$ takes values in a simple,
connected, compact group $G$, and $\tg$ is an extension of $g$ to
$\B$ where $\partial \B=\M$. $\M$ is the two-dimensional space-time,
and we take space to be compactified so that $\M= S^{1} \times \bbbr$.
$\langle .,.\rangle$ is the Killing form on the Lie algebra $\lg$ of
$G$, normalised so that the longest roots of the algebra have length
square equal to $2$.

\noi We shall mainly be interested in the case where $G$ is not
simply-connected. $G$ can then be written as
\be
G = \tG / C \,,
\ee
where $\tG$ is the universal covering group of $G$, and $C$ is a
subgroup of the centre $\Z$ of $\tG$. The centre is isomorphic
to $\Zop_N$ (for some $N$) for all simply-connected simple compact
groups, with the exception of $D_{2 n} \cong SO(4n)$ for which the
centre is $\Zop_2\times\Zop_2$.

\noi The second term in (\ref{action}), the Wess-Zumino term, depends
on the choice of the extension $\tg$, and the consistency of the quantum
theory requires this ambiguity to be $2\pi \Zop$. This
imposes quantisation conditions on $k$. For $C=\Zop_{N}$, the
quantisation condition is \cite{FGK2}
\be
\label{quantis}
\frac{k N}{2} \langle c_1, c_2 \rangle \in \Zop \,,
\ee
for all $c_i\in C$, and for $C=\Zop_2\times\Zop_2$,
the condition is
\be
k \langle c_1, c_1 \rangle, \hspace*{0.5cm}
k \langle c_2, c_2 \rangle,  \hspace*{0.5cm}
k \langle c_1, c_2 \rangle \in \Zop \,,
\ee
where $c_1=(1,0)$ and $c_2=(0,1)$. Here we have identified (as we
shall do from now on) $\hat{c}\in C\subset \Z\subset \tG$ with
$c$ in the quotient space of a Cartan subalgebra $\lh$ by the coroot
lattice via $\hat{c}=exp(c)$. For $C=\Zop_2\times\Zop_2$, the definition of
the Wess-Zumino term remains ambiguous and there are at least two
quantisations \cite{FGK2}.

\noi Recall that the most general solution $g(x,t)$ of the (classical)
WZW-model is of the form
\be
g(x,t) = u(x^+) \cdot v(x^-) \,,
\ee
where $x^{\pm} = t \pm x$ and $u,v:\bbbr \rightarrow G$. The solution
$g$ has to be periodic in $x\mapsto x+2 \pi$, and this implies that
$u$ and $v$ have to satisfy
\be
\label{monodromy}
u(z+2 \pi) = u(z) M, \hspace*{0.5cm} v(z+2 \pi) = M v(z) \,,
\ee
where $M$ is some group element. The pair of functions $(u,v)$ does
not uniquely determine the solution $g$; the transformation
\be
u \rightarrow u g_0 \hspace*{0.5cm} v \rightarrow g_0^{-1} v \,,
\ee
under which
\be
M \rightarrow g_0^{-1} M g_0
\ee
leaves $g$ invariant. We can use this freedom to rotate $M$ into a
fixed maximal torus $H$ of $G$. This fixes the pair of functions
$(u,v)$ generically up to
\be
\label{gauge}
u \rightarrow u n \hspace*{0.5cm} v \rightarrow n^{-1} v \,,
\ee
where $n\in N(H)$, the normaliser of $H$, under which $M\rightarrow
n^{-1} M n$ \cite{MPHOS}. (If $M$ is not regular, {\it i.e.} if $M$
belongs to more than one maximal torus, the residual symmetry is even
larger.)  The action of $n$ on $M$ has kernel $H$; the quotient
$\W=N(H)/H$ is the Weyl group of $G$.
\medskip

\noi The configuration space of the WZW model is the space of
functions $\M\rightarrow G$. We could now attempt to describe the phase
space of the system in terms of coordinates (on this configuration
space) and their conjugate momenta, but this would seem to be rather
difficult. We shall therefore use a different description which was
already employed in \cite{MPHOS} and goes back to \cite{CWZ}. In this
approach, the phase space is regarded as the manifold of solutions of
the equations of motion of the field theory. As in \cite{MPHOS}, this
space can be regarded as a symplectic quotient of a larger space in
which we relax the constraint that the monodromy of $u$ and $v$ should
be the same. That is, we introduce $\nu_L, \nu_R \in \lh$, the Lie
algebra of the maximal torus $H$, and write the left and right
monodromy, respectively, as
\be
M_L= exp(2 \pi \nu_L) \hspace{0.5cm} M_R= exp(2 \pi \nu_R)\,.
\ee
We can then move all the non-trivial homotopic information about
$u$ and $v$ into $\nu_L$ and $\nu_R$, respectively, {\it i.e.} we can
write
\be
u(x^+) = \tu(x^+) exp(\nu_L x^+)
\ee
and
\be v(x^-) = exp(\nu_R x^-) \tv(x^-)\,,
\ee
where $\tu,\tv\in L\tG$, the loop group of the simply connected
covering group $\tG$ of $G$. We can restrict $\nu_L$ and $\nu_R$,
without loss of generality, to lie in a fixed alcove (a chamber of the
Stiefel diagram) of $\lg$ \cite{BtD}. Then the phase space is the
quotient of the submanifold $exp(2 \pi \nu_L) = exp(2 \pi \nu_R)$ (as
a relation in $G$) by the action of $(H\times C)$,
\be
\label{phase}
\cup_{c\in C} \left( \left. (\tu,\nu_L,\nu_R,\tv)
\right|_{\nu_L=c\cdot \nu_R} \right) / (H\times C) \,,
\ee
where $H$ acts on $\tu$ and $\tv$ as in (\ref{gauge}) and leaves
$\nu_L$ and $\nu_R$ invariant, and $c\in C$ maps
$(\tu,\nu_L,\nu_R,\tv)$ to $(\tu,\nu_L,\nu_R,\tv\, c)$. To describe
the action of $c\in C$ on $\nu_R$ we regard (as before) $C\cong
\pi_1(G)$ as the quotient space of the lattice of integral elements of
$\lh$, {\it i.e.} those that are mapped to $1$ in $G$, by the coroot
lattice; this quotient space acts naturally on a fixed alcove by
translation.
\medskip

\noi To quantise the theory we should now find a subalgebra of
functions on phase space which can be consistently defined as
operators in the quantum theory, replacing Poisson brackets by
commutators. Unfortunately, the phase space is rather complicated (in
particular it is not a vector space), and it is therefore very
difficult to find such a subalgebra explicitly. On the other hand, we
might argue that the subalgebra should contain the analogues of the
position and momentum function, and that therefore the phase space
itself (regarded as a subspace of the space of distributions on phase
space) should be contained in the closure of this subalgebra. Then,
the quantum states should form a representation of (\ref{phase}), and
thus of
\be
\label{phasesectors}
\left( \cup_{c\in C} \left. (\tu,\nu_L,\nu_R,\tv)
\right|_{\nu_L=c\cdot \nu_R} \right) \,,
\ee
which is covariant under the action of $(H\times C)$. (This
description resembles strongly the formulation in \cite{MG4}.) For
$G=\tG$, $C$ is the trivial group, and (\ref{phase}) consists of one
component only. In this case the configuration space (and the phase
space) is simply connected, and there should exist only one
quantisation in which the quantum states form a representation of
(\ref{phasesectors}) which is invariant under $H$.  This should force
the representations of $\tu$ and $\tv$ to be conjugate, and we expect
therefore that the spectrum only contains states in the diagonal
theory. Indeed, it is known \cite{FGK2} that the theory for $\tG$ is
\be
\label{diagonal}
\H_{\tG} = \sum_{j} \H_j \otimes \H_{\jbar} \,,
\ee
where the sum extends over all unitary positive energy representations
of $\tL\tG$, the central extension of the loop group $L\tG$
\cite{PS}, and $\H_{\jbar}$ denotes the conjugate representation to $\H_j$.

\noi In the general case, the spectrum of the theory corresponding to
$G=\tG/C$ contains also states which are representations of the other
components of (\ref{phasesectors}). The additional states of the
quantum theory should be obtained by the action of the non-trivial
loops of $G$ (which are labelled by $c\in C$) on one of the two
representations in the tensor products of (\ref{diagonal}). Recall
from Pressley and Segal \cite{PS}, that the action (by conjugation) of
the non-trivial loops of $G$ on $\tL\tG$ is well-defined. This action
does not preserve the set of positive roots, but, as has been shown in
\cite{FGK2} (see also the appendix), there exists a unique element in
the affine Weyl group $\bW(\lg)$, so that the composition with this
affine Weyl group element preserves the positive roots. The
non-trivial loops of $G$ therefore induce (outer) automorphisms of
$\tL\tG$, and thus map in general a representations of $\tL\tG$,
$\H_j$, into a different representation which we denote by
$\H_{c(j)}$. (The map $c(j)$ has been calculated for all simple groups
in \cite{FGK2}.)

\noi Each component of the configuration space (and also of the phase
space (\ref{phase})) is not simply connected, the fundamental group
being isomorphic to $C$. We therefore expect that there should be
different quantisations of the classical theory corresponding to
different monodromies with respect to $C$. \footnote{As $C$ is
abelian, only one-dimensional representations of the fundamental group
appear.} To classify the different quantisations, recall that for any
subgroup $C$ of the centre of $\tG$, the irreducible representations
of $\tG$ (and thus also of $\tL\tG$) fall into equivalence classes
which are characterised by the induced (one-dimensional)
representation of $C$. The tensor product of two irreducible
representations is in the equivalence class corresponding to the
product representation. For each sector, labelled by $c\in C$, let
$R_c$ denote the equivalence class of representations of $\tL\tG$
which corresponds to the representation $R_c: C \rightarrow U(1)$. The
possible quantisations are then of the form
\be
\label{gentheo}
\H_G^{R} : = \sum_{c\in C} \sum_{j\in R_c} \H_j \otimes \H_{c(\jbar)}\,,
\ee
where $R_c$ has to satisfy
\be
\label{Rconsis}
R_{c_1} R_{c_2} = R_{c_1 + c_2} \,,
\ee
so that the theory is closed under operator products. This means
that $R$ must be a homomorphism from $C\cong \pi_1(G)$ to
$\mbox{Hom}(\pi_1(G),U(1))$. For $C\cong \Zop_N$,
\be
\mbox{Hom}(\Zop_N,\mbox{Hom}(\Zop_N,U(1))) \cong
\mbox{Hom}(\Zop_N,U(1))\,,
\ee
as each $R\in \mbox{Hom}(\Zop_N,\mbox{Hom}(\Zop_N,U(1)))$ is already
uniquely determined by $R_{c_1}$, the representation corresponding to
the generator $c_1$ of $\Zop_N$. On the other hand, for
$C\cong\Zop_2\times\Zop_2$,
\be
\mbox{Hom}(\Zop_2\times\Zop_2,\mbox{Hom}(\Zop_2\times\Zop_2,U(1))) \cong
\mbox{Hom}(\Zop_2\times\Zop_2,U(1))\times
\mbox{Hom}(\Zop_2\times\Zop_2,U(1)) \,,
\ee
as every $R\in \mbox{Hom}(\Zop_2\times\Zop_2,
\mbox{Hom}(\Zop_2\times\Zop_2,U(1)))$ is
uniquely determined by $R_{(1,0)}$ and $R_{(0,1)}$.

\noi Thus, there should exist different quantisations of the
WZW model corresponding to $G=\tG/C$, which are labelled by
$\mbox{Hom}(C,U(1))$ (for $C\cong \Zop_N$) and by
$\mbox{Hom}(\Zop_2\times\Zop_2,U(1))\times
\mbox{Hom}(\Zop_2\times\Zop_2,U(1))$ (for $C\cong
\Zop_2\times\Zop_2$).

\noi In general, these theories are not invariant under the monodromy
corresponding to the analytic continuation of one field
about another one. However, as we shall show in the next section, the
operator product transforms covariantly with respect to a
one-dimensional representation. The appearance of an `anyonic'
representation, describing the monodromy, indicates that the
corresponding theories are genuinely braided. The appearance of the
braid group is quite generic for two-dimensional quantum field
theories \cite{FM,Fre,FGR}. On the other hand, such theories have
traditionally been excluded in Euclidean conformal field theory.

\noi The theories are also not modular invariant in general. However,
as we shall show, the modular invariant theory of Felder, Gawedzki and
Kupiainen \cite{FGK2} is one of the possible quantisations. It
corresponds to a quantisation which is invariant under monodromy, but
it is not characterised by this property alone. In fact, we shall show
that for $C=\Zop_N$ with $N$ even, there exists another monodromy
invariant theory (which is then not modular invariant), and for
$C=\Zop_2 \times \Zop_2$, there exist in addition three monodromy
invariant quantisations for each of the two different modular
invariant solutions.

\noi We should also mention that the above analysis resembles somewhat
the treatment in \cite{SY}, where simple currents were used to
construct modular invariant partition functions via an orbifold
construction.
\smallskip

\noi From a general point of view, regarding the quantum states as
wave-functions on configuration space, one would expect that the
different quantisations correspond to the different choices for the
`Aharanov-Bohm' phases in each component of the configuration
space. If we insist that the quantum theory should be symmetric under
the full loop group (which acts naturally on the configuration space),
then the choice for the `Aharanov-Bohm' phases in the identity
component fixes the phases in all other components. We would thus
expect that the different quantisations are parametrised by
$\mbox{Hom}(\pi_1(G),U(1))$ as explained in \cite{MPHOS}. However,
since we are interested in a quantum field theory where states and
fields are in one-to-one correspondence, we have the additional
constraint that the theory should be closed under the operator
product. It is natural to believe that the Aharanov-Bohm phases
multiply under the operator product, and then the closure condition
implies that the phases of the identity component have to be
trivial. We therefore expect that there exists only one quantum field
theory which is symmetric under the full loop group, the theory with
trivial Aharanov-Bohm phases. This theory, however, is not modular
invariant in general as the modular invariant theories do not always
satisfy $R_c=id$ for all $c\in C$ \cite{FGK2}.

\noi The above theories (\ref{gentheo}) are only symmetric under the
identity component of the loop group, and thus, a priori, the phases
in the different components are unrelated. The possible theories are
then selected by the condition that they are closed under operator
products. As the phases multiply, this implies that the different
quantisations are parametrised by
$\mbox{Hom}(\pi_1(G),\mbox{Hom}(\pi_1(G),U(1)))$. Depending on the
structure of the fundamental group one or two of these theories are
the modular invariant theories of \cite{FGK2}. These theories were
selected in \cite{FGK2} by the specific choice \cite[eq.\ (4.9)]{FGK2},
relating the Aharanov-Bohm phases in the different components.

\section{Monodromy covariance}
\renewcommand{\theequation}{3.\arabic{equation}}
\setcounter{equation}{0}

The key step in the analysis of the monodromy is the calculation of
the transformation of the spectrum of $L_0$ under conjugation by a
loop corresponding to $c\in C$. A priori, as described in Pressley and
Segal \cite{PS}, the action (by conjugation) of the non-trivial loops
of $G$ on $\tL\tG$ is well-defined, and thus induces a well-defined
action on the (untwisted) Kac-Moody algebra $\hat{\lg} \cong
l\lg\oplus\Rop$, the Lie algebra of $\tL\tG$. On the other hand, a
priori a choice has to be made for the definition of the conjugation
on $\Rop\oplus l\lg \oplus\Rop$, the Lie algebra of the extension
$\Pi\triangleright\tL\tG$ of $\tL\tG$ by the rigid rotations $\Pi$
whose generator is $L_0$ (see section~4.9 of \cite{PS} and the
appendix). This ambiguity can be removed by identifying $L_0$ with the
Sugawara expression which is quadratic in $\hat{\lg}$.

\noi The Kac-Moody algebra $\hat{\lg}$ can be described in a modified
Cartan-Weyl basis as follows \cite{GO}:
\be
\begin{array}{ccl}
{\displaystyle [H^i_m, H^j_n]} & = &
{\displaystyle k m \delta^{ij} \delta_{m,-n} \vspace*{0.3cm} }  \\
{\displaystyle [H^i_m, E^{\alpha}_n]} & =  &
{\displaystyle \alpha^i E^{\alpha}_{m+n} \vspace*{0.3cm} }  \\
{\displaystyle [E^{\alpha}_m, E^{\beta}_n]} & = &
\left\{
\begin{array}{ll}
{\displaystyle \varepsilon(\alpha,\beta) E^{\alpha+\beta}_{m+n}
\vspace*{0.2cm} } &
\mbox{if $\alpha+\beta$ is a root, \vspace*{0.2cm} } \\
{\displaystyle \frac{2}{\alpha^2} \left( \alpha \cdot H_{m+n} + k m
\delta_{m,-n} \right)} & \mbox{if $\alpha=-\beta$ ,\vspace*{0.2cm} } \\
{\displaystyle 0} & \mbox{otherwise.\vspace*{0.3cm} }
\end{array} \right. \\
{}[k,E^{\alpha}_n] & =  & [k,H^i_n] = 0\,.
\end{array}
\ee
Here $i=1, \ldots, r=\mbox{rank}~\lg$, $\alpha$ labels the positive roots
$R^+$ of $\lg$, and the horizontal subalgebra (with $n=0$) is isomorphic
to $\lg$.
\smallskip

\noi Upon conjugation with the loop $\theta\in [0,2 \pi ] \mapsto
exp(\theta c)$, where $exp(2 \pi c)\in\Z$, the centre of $\tG$, the
generators of the Kac-Moody algebra transform as \cite{PS}
\be
H^i_m \mapsto H^i_m - \delta_{m,0} k \langle c, H^i \rangle \,,
\ee
\be
E^{\alpha}_m \mapsto E^{\alpha}_{m - \langle \alpha,c \rangle} \,,
\ee
\be
k \mapsto k \,,
\ee
where $\langle .,. \rangle$ denotes the Killing-form of the horizontal
subalgebra.
\medskip

\noi The Sugawara expression for $L_0$ is given as \cite{GO}
$$
L_0 = \frac{1}{\beta} \left[
\sum_{i=1}^{r} H^i_0 H^i_0 +
\sum_{\alpha\in R^+} \left( E^{\alpha}_0 E^{-\alpha}_0 +
E^{- \alpha}_0 E^{\alpha}_0 \right) \right. \hspace*{4cm}
$$
\be
\hspace*{3cm}
\left. + 2 \sum_{n=1}^{\infty}
\left( \sum_{i=1}^{r} H^i_{-n} H^i_n +
\sum_{\alpha\in R^+} \left( E^{\alpha}_{-n} E^{-\alpha}_n
+ E^{-\alpha}_{-n} E^{\alpha}_n \right) \right) \right] \,,
\ee
where $\beta=2 k + Q_{\psi}$, and $Q_{\psi}$ is the quadratic Casimir
in the adjoint representation (with highest weight $\psi$, where
$\psi$ is the highest root). Upon conjugation with the loop
$\theta\in [0,2 \pi ] \mapsto exp(\theta c)$ $L_0$ becomes
\begin{eqnarray}
L_0^{\prime} & = & \frac{1}{\beta} \left[
\sum_{i=1}^{r} \left( H^i_0 H^i_0 - 2 k \langle c, H^i \rangle H^i_0
+ k^2 \langle c, H^i \rangle \langle H^i,c \rangle \right)
\right. \nonumber \\
& & + \sum_{\alpha\in R^+} \left( E^{\alpha}_{- \langle \alpha, c\rangle}
E^{-\alpha}_{\langle \alpha, c \rangle} +
E^{-\alpha}_{\langle \alpha, c \rangle}
E^{\alpha}_{-\langle \alpha, c \rangle} \right) \nonumber \\
& & \left.
+ 2 \sum_{n=1}^{\infty} \left( \sum_{i=1}^{r} H^i_{-n} H^i_n
+ \sum_{\alpha\in R^+} \left( E^{\alpha}_{-n- \langle \alpha, c \rangle}
E^{-\alpha}_{n + \langle \alpha, c \rangle}
+ E^{-\alpha}_{-n + \langle \alpha, c \rangle}
E^{\alpha}_{n -  \langle \alpha, c \rangle}\right) \right) \right] \,.
\end{eqnarray}
We have the identities
\be
- \sum_{i=1}^{r} 2\, k \,\langle c, H^i \rangle H^i_0 \; | \lambda \rangle =
- 2 \, k \, \langle c,\lambda \rangle \; | \lambda \rangle \,,
\ee
where $\lambda$ is the weight of the state $|\lambda\rangle$ on which
$L_0^{\prime}$ is evaluated, and
\be
\sum_{i=1}^{r} k^2 \langle c, H^i \rangle \langle H^i,c \rangle =
k^2 \langle c,c \rangle \,.
\ee
Furthermore, we can choose (without loss of generality) the set of
positive roots, $R^+$, such that $\langle \alpha, c \rangle \geq 0$ for
all $\alpha\in R^+$. Then we can rewrite $L_0^{\prime} | \lambda \rangle$
as
$$
L_0  | \lambda \rangle
+ \frac{1}{\beta}\Biggl[- 2 k \langle c,\lambda \rangle
+ k^2 \langle c,c \rangle \Biggr.
\hspace*{10.5cm}
\vspace*{-0.6cm}
$$
\begin{eqnarray}
\hspace*{0.1cm}
& & \left.+ \sum_{\alpha\in R^+}
\left( 2 \sum_{n=1}^{\langle \alpha,c\rangle - 1}
[ E^{-\alpha}_{-n + \langle \alpha,c\rangle} ,
E^{\alpha}_{n - \langle \alpha,c\rangle}]
+ [E^{-\alpha}_0, E^{\alpha}_0 ]
+ [ E^{-\alpha}_{\langle \alpha,c\rangle} ,
E^{\alpha}_{- \langle \alpha,c\rangle}] \right) | \lambda \rangle
\right] \nonumber \\
& = & L_0  | \lambda \rangle
+ \frac{1}{\beta}\left[- 2 k \langle c,\lambda \rangle
+ k^2 \langle c,c \rangle
+ 2 \sum_{\alpha\in R^+} \left( \sum_{n=0}^{\langle \alpha,c\rangle -1}
[ E^{-\alpha}_{-n + \langle \alpha,c\rangle} ,
E^{\alpha}_{n - \langle \alpha,c\rangle}]
- \frac{k}{\alpha^2} \langle \alpha,c\rangle \right)  | \lambda \rangle
\right] \nonumber \\
& = & L_0  | \lambda \rangle
+ \frac{1}{\beta}\Biggl[- 2 k \langle c,\lambda \rangle
+ k^2 \langle c,c \rangle  \Biggr. \nonumber \\
& & \left. + \sum_{\alpha\in R^+} \left(
\sum_{n=0}^{\langle \alpha,c\rangle -1}
\left\{ - \frac{4}{\alpha^2} \langle \alpha, \lambda \rangle
+ \frac{4}{\alpha^2} k (-n + \langle \alpha, c \rangle) \right\}
- \frac{2 k}{\alpha^2} \langle \alpha, c \rangle \right) \right] \nonumber \\
& = & L_0 | \lambda \rangle
+ \frac{1}{\beta}\left[- 2 k \langle c,\lambda \rangle
+ k^2 \langle c,c \rangle
+ \sum_{\alpha\in R^+} \left\{
- \frac{4}{\alpha^2} \langle \alpha, \lambda \rangle
\langle \alpha, c \rangle
+ \frac{4}{\alpha^2} k \left( \sum_{l=1}^{\langle \alpha,c\rangle} l\right)
- \frac{2 k}{\alpha^2} \langle \alpha, c \rangle \right) \right] \nonumber \\
& = & L_0 | \lambda \rangle
+ \frac{1}{\beta}\left[- 2 k \langle c,\lambda \rangle
+ k^2 \langle c,c \rangle
+ \sum_{\alpha\in R^+} \left\{
- \frac{4}{\alpha^2} \langle \alpha, \lambda \rangle
\langle \alpha, c \rangle
+ \frac{2 k}{\alpha^2} \langle \alpha,c\rangle \langle c, \alpha\rangle
\right\} \right]\,. \nonumber \\
\end{eqnarray}
Finally, we can use the identity (see {\it e.g.} \cite{GOS})
\be
\sum_{\alpha\in R^+} \frac{4}{\alpha^2}\;\; |\alpha \rangle \langle
\alpha| =  Q_{\psi} \bbbone_r \,,
\ee
where $\bbbone_r$ is the unit matrix in the space of rank~$\lg$
matrices, to conclude that
\be
\label{difference}
L_0^{\prime} \; | \lambda \rangle =
L_0 \; | \lambda \rangle - \langle c,\lambda \rangle \; | \lambda \rangle
+ \half \, k \, \langle c,c \rangle  \; | \lambda \rangle  \,,
\ee
as $\beta = 2 k + Q_{\psi}$.
\medskip

\noi If $c$ is a coroot, the conjugation corresponds to a
transformation in the affine Weyl group $\bW(\lg)$, and the induced
transformation on weights coincides with the formula given in
\cite{GO}. We should also mention that essentially this formula has
been derived in \cite{FGV} for $\lg=a_n$, and, in a different context,
in \cite{Halpern}.
\bigskip

\noi We are now in the position to analyse the monodromy properties of
the different quantisations (\ref{gentheo}). Recall that for any
weight $\lambdabar$ in a subrepresentation of the tensor product of
two representations with highest weight $\lambdabar_1$ and
$\lambdabar_2$, respectively, and for any $exp(2\pi c)\in\Z$, we have
\be
\langle c, \lambdabar \rangle = \langle c, \lambdabar_1 \rangle
+ \langle c, \lambdabar_2 \rangle \;\; (\mbox{mod $1$}) \,.
\ee
Furthermore, the product of two states $(\lambda_1,c_1(\lambdabar_1))$
and $(\lambda_2,c_2(\lambdabar_2))$ in (\ref{gentheo}) is in the
sector $(\lambda,(c_1 + c_2)(\lambdabar))$.  As the $L_0$ spectrum is
the same for a representation and its conjugate, upon rotating the
field corresponding to $(\lambda_1,c_1(\lambdabar_1))$ by $2 \pi$
about $(\lambda_2,c_2(\lambdabar_2))$, the three-point function
changes by
\be
\label{braid}
R(\lambda_1, c_1; \lambda_2, c_2) =
exp \Bigl\{ 2 \pi \Bigl( - \langle \lambdabar_1, c_2 \rangle
- \langle \lambdabar_2, c_1 \rangle +  k \langle c_1, c_2 \rangle \Bigr)
\Bigr\} \,,
\ee
which is independent of $\lambda$. This demonstrates that the fields
of (\ref{gentheo}) are covariant under monodromy with respect to a
one-dimensional representation which is given by (\ref{braid}).

\section{Additional monodromy invariant theories}
\renewcommand{\theequation}{4.\arabic{equation}}
\setcounter{equation}{0}

We want to analyse now, how many monodromy invariant quantisations
exist. Recall from section~2 that the different quantisations are of
the form
\be
\label{localtheo}
\H_G^R := \sum_{c\in C} \sum_{j\in R_c} \H_j \otimes \H_{c(\jbar)} \,,
\ee
where $G=\tG/C$ and $R_c$ is an equivalence class of positive energy
representations of $\hat{\lg}$ corresponding to a representation of
$C$ (which we also denote by $R_c: C \rightarrow U(1)$).
In order for the theory to be closed under composition, the
assignment of $R_c$ to $c$ must respect the group structure of $C$,
{\it i.e.} $R$ must be an element of $\mbox{Hom}(C,\mbox{Hom}(C,U(1)))$.
The quantisation is invariant under monodromy, if $R(\lambda_1,
c_1;\lambda_2, c_2)=1$ for all $(\lambda_i,c_i(\lambdabar_i))$ in
(\ref{localtheo}). Let us consider the two different cases for the
structure of $C$ separately.

\subsection{The case $C=\Zop_N$.}

Let $c$ denote a cyclic generator of $C$, where the order of $c$ is
$N$. Because of the representation property of $R$, all
$R_{c^\prime}$ are uniquely determined, once $R_c$ is fixed. Let
$\lambda\in R_c$. By requiring monodromy invariance for the product of
$(\lambda,c)$ with itself, we find using (\ref{braid})
\be
0 = 2 \left( - \langle \lambdabar,c\rangle + \half k \langle c,c\rangle
\right) \;\; (\mbox{mod $1$})\,.
\ee
As $C$ has order $N$, any possible weight $\lambdabar$ has to satisfy
\be
\label{constr}
N \langle \lambdabar,c\rangle = 0 \;\; (\mbox{mod $1$}) \,,
\ee
and any linear functional, satisfying (\ref{constr}), corresponds to a
class of possible weights. The quantisation condition
\be
\label{quanti}
\half N k \langle c,c\rangle \in \Zop
\ee
thus guarantees that there exists $\lambdabar$ such that
\be
- \langle \lambdabar,c\rangle + \half k \langle c,c\rangle =0 \;\;
(\mbox{mod $1$})\,.
\ee
In this case, the difference of the $L_0$ eigenvalues of the
representation corresponding to $\lambda$ and to $c(\lambdabar)$ is
integral because of (\ref{difference}) and since a representation and
its conjugate have the same $L_0$ spectrum. The same is true
for all other sectors, since
\be
- \langle m \lambdabar, m c \rangle + \half k \langle m c, m c \rangle
= m^2 \left( - \langle \lambdabar,c\rangle + \half k \langle c,c\rangle
\right) = 0 \;\; (\mbox{mod $1$}) \,.
\ee
Thus this solution corresponds to the unique modular invariant theory
of Felder {\it et.al.} \cite{FGK2}.
\medskip

\noi Because of the consistency condition (\ref{constr}) and the
quantisation condition (\ref{quanti}), there exists a solution
$\lambdabar_1$, satisfying
\be
- \langle \lambdabar_1 , c \rangle + \half k \langle c, c \rangle =
\half\;\; (\mbox{mod $1$})\,,
\ee
only if $N$ is even. On the other hand, for even $N$, $\lambdabar_1$
is a possible weight as it corresponds to a representation in the
equivalence class of the one-dimensional representation of the centre
\be
R(w) = - e^{\pi i k \langle w,w \rangle} \,,
\ee
where $w\in C$. To check that the corresponding theory is monodromy
invariant, we observe that
\be
- \langle m \lambdabar_1, n c \rangle
- \langle n \lambdabar_1, m c \rangle
+ k \langle m c, n c \rangle
= m n \left( - 2 \langle \lambdabar_1, c \rangle
+ k \langle c, c \rangle \right) \in \Zop \,.
\ee
Thus all sectors are relatively monodromy invariant. It is clear
that this additional monodromy invariant theory is not modular
invariant, as the partition function is not invariant under
$\tau \mapsto \tau + 1$.

\subsection{The case $C=\Zop_2\times\Zop_2$.}

Let $c_1 = (1,0)$ and $c_2=(0,1)$ be the two generators of
$C=\Zop_2\times\Zop_2$, and denote by $\lambda_1$ and $\lambda_2$ the
corresponding weights in (\ref{localtheo}). We have to check, case by
case, the conditions implied by monodromy invariance..

\begin{list}{(\roman{enumi})}{\usecounter{enumi}}
\item $(1,0) \times (1,0) = (0,0)$. The requirement is
\be
- 2 \langle \lambdabar_1, c_1 \rangle + k \langle c_1, c_1 \rangle =
0 \;\; (\mbox{mod $1$}) \,.
\ee
As $2c_1$ is a coroot, $2 \langle \lambdabar_1, c_1 \rangle \in\Zop$, and
the condition is
\be
k \langle c_1, c_1 \rangle \in \Zop \,,
\ee
which is one of the quantisation conditions.
\item $(0,1) \times (0,1) = (0,0)$. An identical reasoning gives
\be
k \langle c_2, c_2 \rangle \in \Zop \,,
\ee
which is again one of the quantisation conditions.
\item $(1,1) \times (1,1) = (0,0)$. Similarly we find
\be
k \langle c_1+c_2, c_1+c_2 \rangle \in \Zop \,,
\ee
which follows from one of the quantisation conditions.
\item $(1,0) \times (0,1) = (1,1)$. Using (\ref{braid}), we find
\be
- \langle \lambdabar_1, c_2 \rangle - \langle \lambdabar_2, c_1 \rangle
+ k \langle c_1, c_2 \rangle \in \Zop \,.
\ee
\end{list}
As explained in \cite{FGK2}, there are two modular invariant theories.
One is characterised by
\be
\label{sol1}
\begin{array}{lcl}
{\displaystyle
\langle \lambdabar_1,c_1 \rangle} & = &
{\displaystyle \frac{k}{2} \langle c_1, c_1
\rangle  + \frac{n}{2} \;\; (\mbox{mod $1$})} \vspace*{0.2cm} \\
{\displaystyle \langle \lambdabar_1,c_2 \rangle} & = &
{\displaystyle \frac{k}{2} \langle c_1, c_2
\rangle \;\;(\mbox{mod $1$})}  \vspace*{0.2cm} \\
{\displaystyle \langle \lambdabar_2,c_1 \rangle} & = &
{\displaystyle \frac{k}{2} \langle c_2, c_1
\rangle \;\;(\mbox{mod $1$})} \vspace*{0.2cm} \\
{\displaystyle \langle \lambdabar_2,c_2 \rangle} & = &
{\displaystyle \frac{k}{2} \langle c_2, c_2
\rangle + \frac{m}{2} \;\;(\mbox{mod $1$})  \,,}
\end{array}
\ee
where $m=n=0$, and the other is characterised by
\be
\label{sol2}
\begin{array}{lcl}
{\displaystyle \langle \lambdabar_1,c_1 \rangle} & = &
{\displaystyle \frac{k}{2} \langle c_1, c_1
\rangle + \frac{n}{2} \;\;(\mbox{mod $1$})} \vspace*{0.2cm} \\
{\displaystyle \langle \lambdabar_1,c_2 \rangle} & = &
{\displaystyle \frac{k}{2} \langle c_1, c_2
\rangle +\half \;\;(\mbox{mod $1$})} \vspace*{0.2cm} \\
{\displaystyle \langle \lambdabar_2,c_1 \rangle} & = &
{\displaystyle \frac{k}{2} \langle c_2, c_1
\rangle +\half \;\;(\mbox{mod $1$})} \vspace*{0.2cm} \\
{\displaystyle \langle \lambdabar_2,c_2 \rangle} & = &
{\displaystyle \frac{k}{2} \langle c_2, c_2
\rangle +\frac{m}{2} \;\;(\mbox{mod $1$})  \,,}
\end{array}
\ee
where, again, $m=n=0$.

\noi To each of the two solutions, there exist additional monodromy
invariant solutions given by (\ref{sol1}) and (\ref{sol2}),
respectively, with $(n=1, m=0)$, $(n=0,m=1)$ and $(n=1,m=1)$. It is
immediate that they satisfy all constraints. It is clear that these
theories are not modular invariant, as the partition functions are not
invariant under $\tau \mapsto \tau + 1$.

\section{Examples}
\renewcommand{\theequation}{5.\arabic{equation}}
\setcounter{equation}{0}

In this section we shall give a few non-trivial examples, exhibiting
the additional (monodromy invariant) quantisations. The simplest
example occurs for $G=SO(3)$ and was already pointed out in
\cite{MPHOS}.

\subsection{$G=SO(3)$.}

The first homotopy group of $SO(3)$ is $\pi_1(SO(3))=\Zop_2$, and
the generator of $\pi_1(SO(3))$, written as an element of $\lh$, is
$c=\sqrt{2}/2$. (Recall, that $E^\pm$ are $\pm\sqrt{2}$ in this
notation.) The quantisation condition requires $k$ to be even, as
$\langle c, c \rangle = 1/2$ (see (\ref{quantis})).

\noi The outer automorphism corresponding to $c$ acts on
representations by mapping the representation with spin $j$ to
the one with spin $c(j)=k/2-j$. There are two cases to consider
\begin{list}{(\roman{enumi})}{\usecounter{enumi}}
\item $k=4n$. The modular invariant theory is given by
\begin{eqnarray}
\H_{0} & = &
\Bigl[ \Bigl([0]\otimes [0]\Bigr)
       \oplus \Bigl[1]\otimes [1]\Bigr)
       \oplus \Bigl[2]\otimes [2]\Bigr) \oplus \ldots \Bigr]
\nonumber \\
& & \oplus
\Bigl[ \Bigl([0]\otimes [k/2]\Bigr)
       \oplus \Bigl([1]\otimes [k/2-1]\Bigr) \oplus \ldots \Bigr]
\nonumber \,.
\end{eqnarray}
\item $k=2n+2$. The modular invariant theory is given by
\begin{eqnarray}
\H_{1} & = &
\Bigl[ \Bigl([0]\otimes [0]\Bigr)
       \oplus \Bigl([1]\otimes [1]\Bigr)
       \oplus \Bigl([2]\otimes [2]\Bigr) \oplus \ldots \Bigr]
\nonumber \\
& & \oplus
\Bigl[ \Bigl([1/2]\otimes [k/2-1/2]\Bigr)
       \oplus \Bigl([3/2]\otimes [k/2-3/2]\Bigr) \oplus \ldots \Bigr]
\nonumber \,.
\end{eqnarray}
\end{list}

\noi The additional quantisation which is also monodromy
invariant is for $k=4n$, $\H_1$, and for $k=2n+2$, $\H_0$. To check
that these theories are indeed monodromy invariant, we note that the
conformal weight of the lowest energy space of the representation
$[j]$ is
\be
L_0([j]) = \frac{j(j+1)}{k+2} \,,
\ee
and that
\be
L_0([c(j)]) - L_0([j]) = \frac{k}{4} - j\,.
\ee
This agrees with (\ref{difference}), as $\langle c, j \rangle = j$,
and the claimed monodromy invariance is easily verified.

\subsection{Quotient groups of $G=SU(4)$.}

\noi The centre of $SU(4)$ is $\Z\cong \Zop_4$, and the group of
central elements is generated by $c=\lambda_1$, the
fundamental weight corresponding to the first root (for the notation see
for example \cite{Gourdin}). The central
elements act (as outer automorphisms) on the representations
$[l,m,n]$, written in the Dynkin basis, as
\be
\begin{array}{lcl}
c([l,m,n]) & = &
{\displaystyle [m,n,k-l-m-n]\,,} \\
c^2([l,m,n]) & = &
{\displaystyle [n,k-l-m-n,l]\,,} \\
c^3([l,m,n]) & = &
{\displaystyle [k-l-m-n,l,m]\,.}
\end{array}
\ee
The conformal weight of the lowest energy space of the representation
$[l,m,n]$ is
\be
\label{local2}
L_0([l,m,n]) = \Bigl( \frac{1}{4} (3 l^2 + 4 m^2 + 3 n^2 + 4 lm + 4mn +
2ln) + (3l + 4m + 3n) \Bigr) / (2(k+4)) \,.
\ee
Using these explicit formulae, it is easy to check that
(\ref{difference}) holds.
\medskip

\noi The centre contains the two different subgroups $\Zop_2$ and
$\Zop_4$, and thus there are two different quotient groups whose
simply connected covering group is $SU(4)$. Let us discuss the two
cases in turn.

\begin{list}{(\roman{enumi})}{\usecounter{enumi}}
\item $G=SU(4)/ \Zop_4$. The quantisation condition is $k\in 2\Zop$,
as $\langle c, c \rangle= 3/4$ and $N=4$ (see (\ref{quantis})).
We have to consider the four cases $k=8p+2s$, where $p\in\Zop$ and
$s=0,1,2,3$.
\begin{itemize}
\item $k=8p$. The modular invariant partition function is
\begin{eqnarray}
\H_0 & = &
\Bigl[ \Bigl([0,0,0]\otimes [0,0,0]\Bigr)
       \oplus \Bigl([1,0,1]\otimes [1,0,1]\Bigr) \oplus \ldots
\Bigr] \nonumber \\
& & \oplus
\Bigl[ \Bigl([0,0,0]\otimes [0,0,k]\Bigr)
       \oplus \Bigl([1,0,1]\otimes [0,1,k-2]\Bigr) \oplus \ldots
\Bigr]
\nonumber \\
& & \oplus
\Bigl[ \Bigl([0,0,0]\otimes [0,k,0]\Bigr)
       \oplus \Bigl([1,0,1]\otimes [1,k-2,1]\Bigr) \oplus \ldots
\Bigr]
\nonumber \\
& & \oplus
\Bigl[ \Bigl([0,0,0]\otimes [k,0,0]\Bigr)
       \oplus \Bigl([1,0,1]\otimes [k-2,1,0]\Bigr) \oplus \ldots
\Bigr] \,. \nonumber
\end{eqnarray}
\item $k=8p+2$. The modular invariant partition function is
\begin{eqnarray}
\H_1 & = &
\Bigl[ \Bigl([0,0,0]\otimes [0,0,0]\Bigr)
       \oplus \Bigl([1,0,1]\otimes [1,0,1]\Bigr) \oplus \ldots \Bigr]
\nonumber \\
& & \oplus
\Bigl[ \Bigl([1,0,0]\otimes [0,0,k-1]\Bigr)
       \oplus \Bigl([0,1,1]\otimes [1,1,k-2]\Bigr) \oplus \ldots
\Bigr] \nonumber \\
& & \oplus
\Bigl[ \Bigl([0,1,0]\otimes [0,k-1,0]\Bigr)
       \oplus \Bigl([1,1,1]\otimes [1,k-3,1]\Bigr) \oplus \ldots
\Bigr]
\nonumber \\
& & \oplus
\Bigl[ \Bigl([0,0,1]\otimes [k-1,0,0]\Bigr)
       \oplus \Bigl([1,1,0]\otimes [k-2,1,1]\Bigr) \oplus \ldots
\Bigr] \,. \nonumber
\end{eqnarray}
\item $k=8p+4$. The modular invariant partition function is
\begin{eqnarray}
\H_2 & = &
\Bigl[ \Bigl([0,0,0]\otimes [0,0,0]\Bigr)
       \oplus \Bigl([1,0,1]\otimes [1,0,1]\Bigr) \oplus \ldots \Bigr]
\nonumber \\
& & \oplus
\Bigl[ \Bigl([0,1,0]\otimes [1,0,k-1]\Bigr)
       \oplus \Bigl([2,0,0]\otimes [0,0,k-2]\Bigr) \oplus \ldots
\Bigr]
\nonumber \\
& & \oplus
\Bigl[ \Bigl([0,0,0]\otimes [0,k,0]\Bigr)
       \oplus \Bigl([1,0,1]\otimes [1,k-2,1]\Bigr) \oplus \ldots
\Bigr]
\nonumber \\
& & \oplus
\Bigl[ \Bigl([0,1,0]\otimes [k-1,0,1]\Bigr)
       \oplus \Bigl([0,0,2]\otimes [k-2,0,0]\Bigr) \oplus \ldots
\Bigr] \,. \nonumber
\end{eqnarray}
\item $k=8p+6$. The modular invariant partition function is
\begin{eqnarray}
\H_3 & = &
\Bigl[ \Bigl([0,0,0]\otimes [0,0,0]\Bigr)
       \oplus \Bigl([1,0,1]\otimes [1,0,1]\Bigr) \oplus \ldots \Bigr]
\nonumber \\
& & \oplus
\Bigl[ \Bigl([0,0,1]\otimes [0,1,k-1]\Bigr)
       \oplus \Bigl([1,1,0]\otimes [1,0,k-2]\Bigr) \oplus \ldots
\Bigr]
\nonumber \\
& & \oplus
\Bigl[ \Bigl([0,1,0]\otimes [0,k-1,0]\Bigr)
       \oplus \Bigl([1,1,1]\otimes [1,k-3,1]\Bigr) \oplus \ldots
\Bigr]
\nonumber \\
& & \oplus
\Bigl[ \Bigl([1,0,0]\otimes [k-1,1,0]\Bigr)
       \oplus \Bigl([0,1,1]\otimes [k-2,0,1]\Bigr) \oplus \ldots
\Bigr] \,. \nonumber
\end{eqnarray}
\end{itemize}

\noi The different quantisations are for each even $k$,
$\H_0, \ldots ,\H_3$. The additional monodromy invariant quantisation
is, for $k=8p$ $\H_2$, for $k=8p+2$ $\H_3$, for $k=8p+4$ $\H_0$, and
for $k=8p+6$ $\H_1$.  Using (\ref{local2}), it is easy to see that
these theories are indeed monodromy invariant.

\item $G=SU(4)/ \Zop_2$. The generator of $\Zop_2$ is $d=2c$.
The quantisation condition is $k\in\Zop$, as $\langle d,d \rangle = 4$
and $N=2$ (see (\ref{quantis})).  We have to consider two cases, $k$
even and $k$ odd.
\begin{itemize}
\item For $k$ even, the modular invariant theory is given by
\begin{eqnarray}
\H_0 & = &
\Bigl[ \Bigl([0,0,0]\otimes [0,0,0]\Bigr)
       \oplus \Bigl([0,1,0]\otimes [0,1,0]\Bigr)  \oplus \ldots \Bigr]
\nonumber \\
& & \oplus
\Bigl[ \Bigl([0,0,0]\otimes [0,k,0]\Bigr)
       \oplus \Bigl([0,1,0]\otimes [0,k-1,0]\Bigr)  \oplus \ldots
\Bigr]\,. \nonumber
\end{eqnarray}
\item For odd $k$, the modular invariant theory is given by
\begin{eqnarray}
\H_1 & = &
\Bigl[ \Bigl([0,0,0]\otimes [0,0,0]\Bigr)
       \oplus \Bigl([0,1,0]\otimes [0,1,0]\Bigr) \oplus \ldots
\Bigr] \nonumber \\
& & \oplus
\Bigl[ \Bigl([1,0,0]\otimes [0,k-1,1]\Bigr)
       \oplus \Bigl([0,0,1]\otimes [1,k-1,0]\Bigr)  \oplus \ldots \Bigr]
\,. \nonumber
\end{eqnarray}
\end{itemize}

\noi Again, as before, the additional quantisation which is also
monodromy invariant is for $k$ even $\H_1$, and for $k$ odd $\H_0$.
As before, the monodromy invariance can be easily checked using
(\ref{local2}).
\end{list}

\subsection{$G=SO(8)/\Zop_2\times \Zop_2$.}

The centre of $G=SO(8)$ is $\Z\cong\Zop_2\times \Zop_2$, and it is
generated by $(1,0) = \lambda_1$, $(0,1) = \lambda_3$ and $(1,1) =
\lambda_4$, where $\lambda_i$ is the fundamental weight corresponding
to the $i$th root. The central elements act (as outer automorphisms)
on the representations $[r_1,r_2,r_3,r_4]$, written in the Dynkin
basis, as \be
\begin{array}{lcl}
c_{(1,0)} ([r_1,r_2,r_3,r_4]) & = &
{\displaystyle [k-r_1-2 r_2-r_3-r_4,r_2,r_4,r_3]\,,} \\
c_{(0,1)} ([r_1,r_2,r_3,r_4]) & = &
{\displaystyle [r_4,r_2,k-r_1-2 r_2-r_3-r_4,r_1]\,,} \\
c_{(1,1)} ([r_1,r_2,r_3,r_4]) & = &
{\displaystyle [r_3,r_2,r_1,k-r_1-2 r_2-r_3-r_4]\,.}
\end{array}
\ee
The conformal weight of the lowest energy space of the representation
$[r_1,r_2,r_3,r_4]$ is
$$
L_0([r_1,r_2,r_3,r_4]) = \Bigl( r_1^2 + 2 r_2^2 + r_3^2 + r_4^2
+ 2 r_2 (r_1 + r_3 + r_4) + r_1 r_3 + r_1 r_4 + r_3 r_4 \Bigr.
\hspace*{2cm}
$$
\be
\hspace*{3cm} \Bigl.
+ 6 r_1 + 10 r_2 + 6 r_3 + 6 r_4 \Bigr) / \left(2 (k+6)\right)\,.
\ee
It is easy to see that the formula for the difference of the
$L_0$ spectrum (\ref{difference}) holds.
\medskip

\noi The quantisation condition is $k\in 2 \Zop$, and for each allowed
$k$ there are two different modular invariant theories. They are
explicitly given as
\be
\begin{array}{lcl}
\H_0 & = & \oplus
{\displaystyle \Bigl[ \left. \Bigl( [r_1,r_2,r_3,r_4] \otimes
[r_1,r_2,r_3,r_4] \Bigr) \right| r_{13}\;\mbox{even};
r_{14} \;\mbox{even}; r_{34} \;\mbox{even}
\Bigr]} \vspace*{0.2cm} \\
& & \oplus
{\displaystyle \Bigl[ \left. \Bigl( [r_1,r_2,r_3,r_4] \otimes
c_{(1,0)}([r_1,r_2,r_3,r_4]) \Bigr) \right|
r_{34}\; \mbox{even}; r_{13}+\frac{k}{2}\;\mbox{even};
r_{14}+\frac{k}{2}\;\mbox{even} \Bigr]} \vspace*{0.2cm} \\
& & \oplus
{\displaystyle \Bigl[ \left. \Bigl( [r_1,r_2,r_3,r_4] \otimes
c_{(0,1)}([r_1,r_2,r_3,r_4]) \Bigr) \right|
r_{14} \; \mbox{even}; r_{13}+\frac{k}{2}\;\mbox{even};
r_{34}+\frac{k}{2} \; \mbox{even} \Bigr]} \vspace*{0.2cm} \\
& & \oplus
{\displaystyle \Bigl[ \left. \Bigl( [r_1,r_2,r_3,r_4] \otimes
c_{(1,1)}([r_1,r_2,r_3,r_4]) \Bigr) \right|
r_{13} \; \mbox{even}; r_{14}+\frac{k}{2}\;\mbox{even};
r_{34}+\frac{k}{2} \;\mbox{even} \Bigr]\,,}
\end{array}
\ee
and
\be
\begin{array}{lcl}
\H_1 & = & \oplus
{\displaystyle \Bigl[ \left. \Bigl( [r_1,r_2,r_3,r_4] \otimes
[r_1,r_2,r_3,r_4] \Bigr) \right| r_{13}\;\mbox{even};
r_{14}\;\mbox{even}; r_{34} \; \mbox{even}
\Bigr]} \vspace*{0.2cm} \\
& & \oplus
{\displaystyle \Bigl[ \left. \Bigl( [r_1,r_2,r_3,r_4] \otimes
c_{(1,0)}([r_1,r_2,r_3,r_4]) \Bigr) \right|
r_{34}\; \mbox{even}; r_{13}+\frac{k}{2}\;\mbox{odd};
r_{14}+\frac{k}{2} \;\mbox{odd} \Bigr]} \vspace*{0.2cm} \\
& & \oplus
{\displaystyle \Bigl[ \left. \Bigl( [r_1,r_2,r_3,r_4] \otimes
c_{(0,1)}([r_1,r_2,r_3,r_4]) \Bigr) \right|
r_{14} \; \mbox{even}; r_{13}+\frac{k}{2}\;\mbox{odd};
r_{34}+\frac{k}{2} \;\mbox{odd} \Bigr]} \vspace*{0.2cm} \\
& & \oplus
{\displaystyle \Bigl[ \left. \Bigl( [r_1,r_2,r_3,r_4] \otimes
c_{(1,1)}([r_1,r_2,r_3,r_4]) \Bigr) \right|
r_{13} \; \mbox{even}; r_{14}+\frac{k}{2}\;\mbox{odd};
r_{34}+\frac{k}{2} \;\mbox{odd} \Bigr]\,,}
\end{array}
\ee
where $r_{ij}=r_i + r_j$, and only those representations appear which
are positive energy, {\it i.e.} satisfy $r_1+2 r_2 + r_3 + r_4 \leq
k$. To shorten the notation, let us describe in the following the theories
by $4 \times 3$ matrices with entries $e$ (even) and $o$ (odd), where the
first row corresponds to $(r_{13}, r_{14}, r_{34})$ in the identity sector,
the second row corresponds to $(r_{34}, r_{13}+ k /2, r_{14} + k/2)$ in
the $c_{(1,0)}$ sector, and so on. For example
\be
\H_0 = \left(\matrix{e & e & e \cr
                     e & e & e \cr
                     e & e & e \cr
		     e & e & e \cr}\right) \hspace*{2cm}
\H_1 = \left(\matrix{e & e & e \cr
                     e & o & o \cr
                     e & o & o \cr
		     e & o & o \cr}\right) \,.
\ee
The additional monodromy invariant theories corresponding to
$\H_0$ are then given as
\be
\H_0^{(1)} =  \left(\matrix{e & e & e \cr
                     o & o & e \cr
                     e & e & e \cr
		     o & e & o \cr}\right) \hspace*{2cm}
\H_0^{(2)} =  \left(\matrix{e & e & e \cr
                     e & e & e \cr
                     o & o & e \cr
		     o & o & e \cr}\right) \hspace*{2cm}
\H_0^{(3)} = \left(\matrix{e & e & e \cr
                     o & o & e \cr
                     o & o & e \cr
		     e & o & o \cr}\right) \,,
\ee
and the additional monodromy invariant theories
corresponding to $\H_1$ are
\be
\H_1^{(1)} =  \left(\matrix{e & e & e \cr
                     o & e & o \cr
                     e & o & o \cr
		     o & o & e \cr}\right) \hspace*{2cm}
\H_1^{(2)} =  \left(\matrix{e & e & e \cr
                     e & o & o \cr
                     o & e & o \cr
		     o & e & o \cr}\right) \hspace*{2cm}
\H_1^{(3)} = \left(\matrix{e & e & e \cr
                     o & e & o \cr
                     o & e & o \cr
		     e & e & e \cr}\right)\,.
\ee
{}From the formulae given above, it is easy to see that these theories
are indeed monodromy invariant. Furthermore, there are another $8$
quantisations which are not monodromy invariant
$$
\H_9 = \left(\matrix{e & e & e \cr
                     e & e & e \cr
                     e & o & o \cr
		     o & e & o \cr}\right) \hspace*{0.8cm}
\H_{10} = \left(\matrix{e & e & e \cr
                     e & e & e \cr
                     o & e & o \cr
		     e & o & o \cr}\right) \hspace*{0.8cm}
\H_{11} = \left(\matrix{e & e & e \cr
                     e & o & o \cr
                     e & e & e \cr
		     o & o & e \cr}\right) \hspace*{0.8cm}
\H_{12} = \left(\matrix{e & e & e \cr
                     e & o & o \cr
                     o & o & e \cr
		     e & e & e \cr}\right) \,,
$$
$$
\H_{13} = \left(\matrix{e & e & e \cr
                     o & e & o \cr
                     e & e & e \cr
		     e & o & o \cr}\right) \hspace*{0.8cm}
\H_{14} = \left(\matrix{e & e & e \cr
                     o & e & o \cr
                     o & o & e \cr
		     o & e & o \cr}\right) \hspace*{0.8cm}
\H_{15} = \left(\matrix{e & e & e \cr
                     o & o & e \cr
                     o & e & o \cr
		     o & o & e \cr}\right) \hspace*{0.8cm}
\H_{16} = \left(\matrix{e & e & e \cr
                     o & o & e \cr
                     e & o & o \cr
		     e & e & e \cr}\right) \,.
$$

\section{Conclusions}

We have analysed the different quantisations of the WZW model
corresponding to a in general non-simply connected group $G$, and we
have found that they are parametrised by
$\mbox{Hom}(\pi_1(G),\mbox{Hom}(\pi_1(G),U(1)))$. In general the
different quantisations are genuinely braided and neither monodromy
nor modular invariant. However, all the modular invariant theories of
Felder {\it et.al.} \cite{FGK2} are contained among them. Furthermore,
for $\pi_1(G)\cong \Zop_N$ with $N$ even, there is another monodromy
invariant theory, and for $\pi_1(G)\cong \Zop_2\times\Zop_2$ there are
another $6$ monodromy invariant quantisations.

\noi It might be hoped that the different quantisations of the WZW
model are characterised by the property to be {\it modular
covariant}. This would mean that it should be possible to define
amplitudes for these theories on arbitrary Riemann surfaces. These
amplitudes would not be invariant under the action of the modular
group. However, they would transform with respect to a
(one-dimensional) representation, and therefore the corresponding
probabilities would be invariant. (The amplitudes would be rather
similar to the amplitudes of a chiral theory, but here they would
correspond to the whole theory.) A first indication that this might be
the case is the fact that the amplitudes transform with respect to a
one-dimensional representation of the monodromy group, as shown in
section~3.
\smallskip

\noi Braided conformal field theories similar to the ones discussed in
this paper have already appeared in \cite{Anni}.\footnote{I thank G.
Watts for drawing my attention to this reference.} There the conformal
limit (in the sector $w=0$) of the $\phi_{3,1}$ (integrable)
perturbation of $\M_{3,5}$ was found to be the theory
\be
\H_{3,5}^{3,1} = \left(\H_{1,1}\otimes\H_{1,1}\right)
\oplus \left( \H_{3,1}\otimes\H_{3,1}\right)
\oplus \left( \H_{2,1}\otimes\H_{3,1}\right)
\oplus \left( \H_{4,1}\otimes\H_{1,1}\right)
\ee
which is genuinely braided. (Our notation follows, for example,
\cite{ID}.) In this context our analysis seems to suggest that the
conformal limit of the different massive perturbations of a conformal
field theory correspond to different quantisations and global
structures of the underlying conformal theory. It would be
interesting to check this conjecture by analysing the conformal limit
of massive perturbations of WZW models.

\noi The analysis of the global properties of the WZW models might
also be relevant for a better understanding of the global issues of
(abelian) $T$-duality in WZW models \cite{Kir,AABL,AAL,RV}. In
particular, similar techniques might be used to analyse in which way
the duality transformation depends on the global topological
properties of the target space group. This is currently work in
progress.

\section*{Appendix}

\appendix

In this appendix we want to establish some notation about the affine
Weyl group. This follows essentially \cite{PS}. We shall also give a
simple geometrical proof for the observation of Olive and Turok
\cite{OT} about the symmetries of the affine Dynkin diagram.

\section{The affine Weyl group}
\renewcommand{\theequation}{A.\arabic{equation}}
\setcounter{equation}{0}

\noi To define the affine Weyl group let us consider the semidirect
product of the loop group $LG$ of $G$ by the unit circle $\Pi$
\be
\Pi\triangleright LG \,,
\ee
where $\Pi$ acts on $LG$ by rigidly rotating the loops. The
introduction of $\Pi$ is completely analogous to the introduction of
the additional generator $d=-L_0$ (see {\it e.g.} \cite{GO}) in the
usual description of the affine root system. That is, we can identify
\be
\lambda \in \Pi \longleftrightarrow \lambda^{L_0}\,,
\ee
which acts on $LG$ by rigidly rotating the loop, {\it i.e.}
$\lambda . f(t) = f(\lambda t)$.

\noi A maximal abelian subgroup of $\Pi\triangleright LG$ is
$(\Pi\times H)$, where $H$ is a maximal torus in $G$. We can then define
the affine Weyl group, analogously to the definition of the Weyl group
of $G$, as
\be
\bW(G) = N(\Pi\times H) / (\Pi\times H) \,,
\ee
where $N(\Pi\times H)$ is the normaliser of $(\Pi\times H)$ in
$\Pi\triangleright LG$.

\noi As has been shown in \cite{PS}, $\bW(G)$ is the semidirect
product of the coweight lattice $\check{H}$ of $G$ by $\W$, the
Weyl group of $G$. Here, the coweight lattice is the lattice of all
homomorphisms
\be
\Pi \rightarrow H \,.
\ee
For a simply connected group, the coweight lattice is generated by the
coroots, and thus
\be
\bW(\tG) = \bW(\lg) \,,
\ee
where the right hand side denotes the usual affine Weyl group,
associated to the affine Lie algebra $\hat{\lg}$ (see {\it e.g.}
\cite{GO}). However, in general, the affine Weyl group $\bW(G)$ is
larger and depends on the group $G$, rather than on the algebra alone.

\noi To be more specific, let us calculate the action of a coweight
$\phi:\Pi\rightarrow H$ on the maximal abelian subgroup $(\Pi\times H)$
of $\Pi\triangleright LG$. Denote a typical element of $(\Pi\times H)$
by $(u,h)$. Then, as $\phi\in LG$,
\begin{eqnarray}
(1,\phi(t))^{-1} . (u,h) . (1,\phi(t)) & = &
(1,\phi(t))^{-1} . (u, \phi(ut) h ) \nonumber \\
& = & (u, \phi(u) h)\,.
\end{eqnarray}
(This demonstrates, in particular, that $\check{H}$ is in the
normaliser of $(\Pi\times H)$.) The action induces an action on roots,
where roots are linear maps
\be
(\Pi\times H) \rightarrow \Pi \,.
\ee
They are specified by $(n,\alpha)\in\Zop\times H^*$, where
\be
(n,\alpha) \Bigl( (u,h) \Bigr) = u^n \alpha(h)\,.
\ee
The induced action of $\phi\in\check{H}$ on $(\Pi\times H)$ is
given as
\begin{eqnarray}
\Bigl(\phi (n,\alpha) \Bigr) (u,h) & = &
(n,\alpha) \Bigl( (1,\phi)^{-1} . (u,h) . (1,\phi) \Bigr) \nonumber \\
& = & u^n \alpha(\phi(u)) \alpha(h) \nonumber \\
\label{conjaction}
& = & (n + \alpha(\phi), \alpha) \Bigl( (u,h) \Bigr) \,.
\end{eqnarray}
Here, we have identified $\phi\in\check{H}$ with $\phi\in\lh$ by
$\phi(t=e^{\theta})=exp(\theta \phi)$. Then
\be
\alpha(\phi(t))=exp(\theta \alpha(\phi)) = t^{\alpha(\phi)}\,.
\ee
For coroots, the action in (\ref{conjaction}) agrees with the formula in
\cite{GO}.
\medskip

\noi So far we have only considered the loop group $LG$, and not its
central extension $\tL G$ by $\Pi$. The Lie algebra of this central
extension is the affine Kac-Moody algebra $\hat{\lg}$, defined in
section~3. For simply connected $G$, the consistency condition on $k$
is that it is integral, and in the general case it is
\be
\label{consis}
k \langle c_1, c_2 \rangle \in \Zop \,,
\ee
where $c_i$ are arbitrary coweights \cite[Section~4.6]{PS}. This
condition is stronger than the quantisation conditions of section~2,
and, in particular, it suggests that one should {\it not} regard the
states of the quantum theory as sections in the $U(1)$ bundle $\tL G$
over $LG$. This is also in accordance with the fact that the theories
we are considering here are only invariant under the identity
component of the loop group, whereas theories whose states are
sections in $\tL G$ would be symmetric under the full loop group.

\noi The action of coweights on $\tL \tG$ induces a unique action on
the Kac-Moody algebra $\hat{\lg}$ \cite[Lemma (4.6.5)]{PS}, and we
have given an explicit formula for this action in section~3. We can
also (re)introduce the extension of $\tL \tG$ by the rigid rotations
$\Pi$.  The action of coroots on the Lie algebra of this extension,
$\Rop\oplus l\lg\oplus\Rop$, is then uniquely determined, but the
action of coweights (which are not coroots) involves a choice.  We can
fix this choice (canonically) by identifying $-d=L_0$ with the
Sugawara expression as we have done in section~3.

\section{Symmetries of the affine Dynkin diagram}
\renewcommand{\theequation}{B.\arabic{equation}}
\setcounter{equation}{0}

Let $\Phi$ and $\bPhi$ be the root systems of $\lg$ and $\hat{\lg}$,
respectively, and denote by $\W$ and $\bW$ the corresponding Weyl
groups. (Here $\bW=\bW(\lg)=\bW(\tG)$ is the affine Weyl group
corresponding to the simply connected group.) As the Weyl groups act
transitively and fixed-point free on the set of simple roots, we have
\be
\mbox{aut}\; \Phi / \W \cong \Gamma \hspace*{3cm}
\mbox{aut}\; \bPhi / \bW \cong \bGamma  \,,
\ee
where $\Gamma$ and $\bGamma$ are the symmetry groups of the
corresponding Dynkin diagrams.

\noi Olive and Turok observed some time ago \cite{OT} that
\be
\label{isomor}
\bGamma / \Z \cong \Gamma \,,
\ee
where $\Z$ is the centre of $\tG$. In this appendix we want to give a
simple geometrical proof of this observation. (A different, algebraic
proof has been given in \cite{GOM}.)

\noi Let $C_0$ be a Weyl chamber of $\lg$ whose walls are made up of
$r=$rank~$\lg$ simple roots of $\lg$, and let $C_1$ be the (unique)
alcove of $G$ satisfying $0\in C_1\subset C_0$ \cite{GO,BtD}. (The
additional wall of $C_1$ corresponds to $(-\psi,1)$, where $\psi$ is
the highest root with respect to the simple roots.) In particular, the
origin is a corner of the alcove.

\noi An automorphism of $\Phi$ maps $C_0$ to a Weyl chamber
$C_0^\prime$ of $\lg$, and using the transitivity of the action of the
Weyl group $\W$ on Weyl chambers, there exists an element in the Weyl
group which maps $C_0^\prime$ back to $C_0$. Furthermore, since $\W$
acts fixed point free, this Weyl group element is uniquely determined.
Conversely, every symmetry transformation of $C_0$ induces a
symmetry of the Dynkin diagram, and thus there is a one-to-one
correspondence between elements in $\Gamma$ and symmetry
transformations of $C_0$. We can similarly apply the same argument to
the affine Dynkin diagram, and thus conclude that $\bGamma$ is in
one-to-one correspondence with symmetry transformations of the alcove
$C_1$.

\noi A given symmetry transformation of $C_1$ corresponds to an
element of $\Gamma$, if and only if the origin is a fixed point of the
symmetry transformation. Furthermore, since all affine hyperplanes
intersect at the origin, any symmetry transformation of $C_1$ must map
the origin to a point in $Z\cap C_1$, where $Z$ is the lattice of
central elements (which is mapped to the centre of $G$ under $exp$).

\noi Now, suppose that $\gamma$ is a symmetry of $C_1$ which maps
the origin to $\gamma(0)\neq 0$. We want to show that there exists an
element $\gamma_1$ in the semidirect product of the lattice of central
elements $Z$ by $\W$,
\be
\gamma_1 \in \bW(\tG / \Z) = \W \triangleright Z \,,
\ee
such that $\gamma_1$ is a symmetry of $C_1$ and
$\gamma_1\circ\gamma(0)=0$. To construct $\gamma_1$, we first
translate $C_1$ by $-\gamma(0)$ which is a translation in $Z$ because
of the previous argument. Then we use the reflections in the simple
roots of $\lg$ to map the translated $C_1$ back into $C_0$. It is
clear that this defines a symmetry transformation of $C_1$, and that
$\gamma_1\circ\gamma(0)=0$.

\noi Conversely, for any point $c\in Z\cap C_1$, we can in this way
construct a symmetry transformation of $C_1$, mapping $c$ to $0$. Thus
the symmetry transformations of $C_1$ which do not preserve the affine
root are in one-to-one correspondence with the points in $Z\cap C_0$,
which in turn is in one-to-one correspondence with the centre of
$\tG$. As the construction preserves the respective group structures,
we have thus shown that (\ref{isomor}) holds. The proof also implies
that
\be
\mbox{aut}\; \bPhi / \bW(\tG/\Z) \cong \Gamma \cong \mbox{aut}\; \Phi / \W
\,.
\ee
\bigskip

\noindent {\bf Acknowledgements}

\noi It is a pleasure to thank Peter Goddard for many enlightening
discussions. I also thank Adrian Kent, Anni Koubek, Noah Linden
and G\'erard Watts for useful conversations.

\noi I am grateful to Pembroke College, Cambridge, for a Research
Studentship.

\end{document}